\documentclass{amsart}
\usepackage{amsmath}
\usepackage{amssymb}
\usepackage{amsthm}
\usepackage{tikz-cd} 
\usepackage{hyperref}
\usepackage{natbib}
\title{Possible and Impossible Inferences From Reconstructed Evolutionary Processes using Phylogenies as an Example}
\author{Niklas Hohmann, Utrecht University \\ (N.H.Hohmann@uu.nl)}
\date{\today}
\begin{document}
\maketitle

\begin{abstract}
    Our understanding of past evolutionary change is often based on reconstructions based on incomplete data, raising fundamental questions about the degree to which we can make reliable inferences about past evolutionary processes. This was demonstrated by Louca and Pennell (2020), who showed that each pure-birth process can be generated by an infinite number of birth-death processes. Here, I explore what it means to reconstruct past evolutionary change with three approaches from measure theory, group theory, and homotopy theory to better understand structural constraint and origins of (non)identifiability. As an example, the developed framework is applied to the case of birth-death processes.
\end{abstract}
\tableofcontents
\section{Introduction}
Recently it has been demonstrated that reconstructions of evolutionary dynamics can have identifiability issue, meaning many evolutionary processes result in the same reconstruction \citep{louca2020}. Generalizing this raises a number of closely related questions I will refer to as the \textbf{identifiability problem:}
\begin{itemize}
\item \textbf{Identifiability or non-identifiability}: Is there a 1-1 correspondence between evolutionary processes and their corresponding reconstructed processes?
\item \textbf{Ambiguity:} In the case of non-identifiability: How ambiguous is the reconstruction, i.e. how many evolutionary processes are mapped to the same reconstructed process?
\item \textbf{Dissimilarity}: How similar or dissimilar are the processes mapped to the same evolutionary process, i.e. how much does the ambiguity matter?
\item \textbf{Distinguishability}: Given two evolutionary processes, can we decide whether they can be distinguished based on reconstructed processes?
\item \textbf{Propagation of non-identifiability:} How do the points above affect further inference about evolutionary processes made on the basis of reconstructed processes?
\item \textbf{Structural origins:} What mathematical properties of the models used control identifiability ar ambiguity?
\item \textbf{Mitigation:} What adjustments can be made to reduce identifiability issues?
\end{itemize}
In this manuscript, I examine these questions using three different approaches (measure theory, theory of group actions, and homotopy theory).\\
In section \ref{model}, I introduce the general modeling framework used throughout this manuscript.\\
Based on this, I use a measure theoretic approach to formalize what it means to extract information from evolutionary and reconstructed processes (section \ref{measure}).\\
Then I analyze the identifiability problem using the theory of group actions and their invariants (section \ref{group}). In section \ref{birthdeath}, this approach is applied to the case of birth-death processes and their reconstructed processes.\\
In section \ref{topology}, I use the theory of covering maps and fundamental groups to examine the structural origins of the identifiability problem, which provides insights into potential mitigation measures.\\
For readability, no proofs are given in this manuscript.
\section{Modeling Framework} \label{model}
I model the inference about evolutionary dynamics as a map $\Omega$ between the set of evolutionary processes, reflecting the true evolutionary dynamics, to the reconstructed processes, reflecting the reconstructed evolutionary dynamics  based on available data. The mapping formalizes the loss of information that occurs between past evolutionary processes their partial reconstruction from available data.\\
For this, let $\mathcal E$ be the set of all evolutionary processes, and $E$ an individual evolutionary process. Similarly, let $\mathcal R$ be the set of all reconstructed processes, and $R$ an individual reconstructed process. The information loss map
\begin{equation}
\Omega \colon \mathcal E \to \mathcal R
\end{equation}
assigns an evolutionary process $E$ to the reconstructed process $R=\Omega(E)$ that is reconstructed from the available data on $E$.\\
For simplicity, I assume that $\mathcal R$ contains only relevant reconstructed processes in the sense that for every reconstructed process $R$ there is an evolutionary process $E$ such that $R$ is the reconstruction of $E$, meaning $\Omega(E)=R$. This is equivalent to the surjectivity of $\Omega$, which can be achieved by either restricting it unto its image or by defining $\mathcal R :=\Omega(\mathcal E)$.\\
A fiber of $\Omega$, given by the preimages of a single reconstructed processes $R$
\begin{equation}
\mathbf C_R:=\Omega^{-1}(R)= \{ E \in \mathcal E \mid \Omega(E)= R \}
\end{equation}
contains all evolutionary processes mapped to $R$. They thus generalize the congruence classes sensu \citet{louca2020} to the setting used here.
\section{Extracting Evolutionary Information from Reconstructed Processes} \label{measure}
In this section, I formalize what it means to extract information about evolutionary processes from reconstructed processes. This provides a characterization of the information that can be extracted from $\mathcal E$ based on the properties of $\Omega$.\\

For this, let $\mathcal F$ be a set of some features of interest. Throughout this section, I assume that $\mathcal E$, $\mathcal R$, and $\mathcal F$ are equipped with $\sigma$-algebras $\mathcal A_{\mathcal E}$, $\mathcal A_{\mathcal R}$, and $\mathcal A_{\mathcal F}$. This is the minimum required structure for a probabilistic analysis, and the elements of the $\sigma$-algebras are the sets that can be assigned probabilities.\\
Let
\begin{equation}
f_{\mathcal E}  \colon (\mathcal E, \mathcal A_{\mathcal E}) \to (\mathcal F, \mathcal A_{\mathcal F})
\end{equation}
be a measurable function that extracts information about the features of interest from the evolutionary processes., meaning $f_\mathcal E (E) = F$ should be read as "the evolutionary process $E$ has the feature $F$".  Conversely, the set of evolutionary processes with features $F$ ($\in \mathcal A_{\mathcal F}$) is given by the preimage $f_{\mathcal E}^{-1}(F)$. We are interested in a function $f_\mathcal R$ that extracts the features of interest not from evolutionary processes, but from reconstructed processes. This motivates the following definition: 
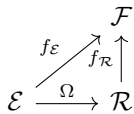
\begin{figure}
    \centering
\begin{tikzcd}
& \mathcal F \\
\mathcal E \arrow[ru, "f_{\mathcal E}"] \arrow[r, "\Omega"] & \mathcal R \arrow[u, "f_{\mathcal R}" ] 
\end{tikzcd}
\caption{Extracting features from evolutionary and reconstructed processes. $f_\mathcal E$ reconstructs features of interest from the evolutinary processes. If it is feconstructable, there exists a function $f_\mathcal R$ that extracts identical features from reconstructed processes.} \label{fig:extracting}
\end{figure}
I call the function $f_{\mathcal E}$ \textbf{reconstructable} if and only if there is a measurable function
\begin{equation}
f_{\mathcal R} \colon (\mathcal R , \mathcal A_{\mathcal R}) \to (\mathcal F , \mathcal A_{\mathcal F} )
\end{equation} 
that satisfies
\begin{equation}
f_{\mathcal E}=  f_{\mathcal R} \circ \Omega
\end{equation}
(see figure \ref{fig:extracting}). Reconstructable functions are exactly the functions that (1) permit a probabilistic analysis and (2) extract information from evolutionary processes that can also be extracted from reconstructed processes. The extraction of information from $\mathcal R$ is performed by $f_{\mathcal R}$.\\
If $\mathcal F$ is a Borel space, reconstructable functions are uniquely characterized by the following statement:
\begin{itemize}
\item A function $f_{\mathcal E}$ is reconstructable if and only if it is $\Omega$-measurable \citep{kallenberg2001}
\end{itemize}
Taking the $\sigma$-algebra $\sigma(\Omega)$ as information system shows that it contains exactly the events $A \subset \mathcal E$ for which the question "Is $E \in A$?" can be decided based on knowledge of $\mathcal R$. Accordingly $\sigma(\Omega)$ contains the maximum information on $\mathcal E$ that can be inferred from $\mathcal R$.\\
Immediate implications of the characterization of reconstructable functions are that the identifiability problem is solvable if and only if the fibers are of size one. Reconstructable functions are constant on congruence classes, so only properties that are shared by all elements of a congruence class can be extracted from the reconstructed processes, and evolutionary processes from the same congruence class can not be distinguished based on reconstructed processes.
\section{Group-Theoretic Approach to the Identifiability Problem} \label{group}
Here I provide a group theoretic approach to the identifiability problem that provides insights into the size of congruence classes, sufficient criteria for distinguishability of evolutionary processes based on reconstructions, and an assessment of information extracted from reconstructed processes.
\subsection{Theoretical Framework}
Throughout this section, I take both evolutionary and reconstructed processes as processes through time, but make no assumption on their state space. For this, let $I$ be the (potentially unbounded) time interval of interest, and assume all involved processes are defined on $I$. Mathematically, this makes evolutionary processes elements of  a function  space $S_E^{\otimes I}$, where $S_E$ is an arbitrary state space of the evolutionary processes. Similarly, reconstructed processes are elements of a function space $S_R^{\otimes I}$, where $S_R$  is an arbitrary state space for the reconstructed processes. The nature of both $S_R$ and $S_E$ are not of interest here, as I will only use transformations of the time interval $I$ for the analysis. \\
A natural operation on processes through time is a time change, meaning a change in the pace with which a process progresses through time. A time change of a process $P$ results in a new process $P'$ that depends both on the time change performed and how the type of process reacts to the time change. For example, rates through time (e.g. extinction rates) will react different to a time change than a accumulated rate (e.g. accumulated extinction rate) or ratios associated with time stamps (e.g. isotope ratios). \\
I formalize the idea of time change in a way that encompasses all these cases. To start, define a time scale $T$ as a strictly increasing, bijective, and continuously differentiable function from $I$ to $I$. It associates points in time  $t_{0}$ with points in time $T(t_0)$ measured in the new time scale $T$. The set of all time scales $\mathcal T$ forms a group with the composition of functions as group operation, the inverse function $T^{-1}$ as inverse element of $T$, and identity $T=\operatorname{Id}$. This reflects that different time changes can be performed successively and can be reversed.\\
Now, write $E_T$ and $R_T$ to indicate that $E$ resp. $R$ are associated with the time scale $T$. If $E$ or $R$ progress through time unchanged, they are associated with the identity time change $\operatorname{Id}$, so $E=E_{\operatorname{Id}}$ and $R=R_{\operatorname{Id}}$.
Based on this, time change can be defined as a group action $\star$ of $\mathcal T$ on 
both $\mathcal E$ and $\mathcal R$ via
\begin{equation}
T \star E_U := E_{T \circ U}
\end{equation}
and
\begin{equation}
T \star R_U := E_{T \circ U}
\end{equation}
The group action of time change naturally defines \textbf{orbits} that consist of all processes that can be transformed into each other via a time change. Mathematically, an orbit $\mathcal O_E$ in $\mathcal E$ is defined as
\begin{equation}
\mathcal O_E := \{E' \in \mathcal E \mid  T \star E=E'\}
\end{equation}
and contains all evolutionary processes that can be generated from $E$ through a change of time scale. Similarly, an orbit $\mathcal O_R$ in $\mathcal R$ is defined as
\begin{equation}
\mathcal O_R := \{R' \in \mathcal R \mid  T \star R=R'\}
\end{equation}
and contains all reconstructed processes that can be generated from $R$ through a change of time scale. For most cases, the processes $E$ or $R$ generating the orbits are not of interest, but I keep  $E$ and $R$ as indices to indicate whether the orbits are subsets of $\mathcal E$ or of $\mathcal R$.
A map $f$ on $\mathcal E$ is called \textbf{invariant} if $f(T \star E) = f(E)$ for all $T \in \mathcal T$ , meaning it is constant on orbits. It is called \textbf{maximalinvariant} if and only if $f(E_1)=f(E_2)$ implies that $E_1$ and $E_2$ are from the same orbit. The same definitions hold mutatis mutandis for maps on $\mathcal R$.\\
I define an \textbf{evolutionary invariant} as a surjective maximalinvariant map
\begin{equation}
\mathcal K_{\mathcal E} \colon \mathcal E \to \mathcal I_{\mathcal E}
\end{equation}
where $\mathcal I_{\mathcal E}$ is some suitable set whose elements uniquely identify the orbits of $\mathcal E$. 
Similarly, I define a \textbf{reconstruction invariant} as a surjective maximalinvariant map
\begin{equation}
\mathcal K_{\mathcal R} \colon \mathcal R \to \mathcal I_{\mathcal R}
\end{equation}
where $\mathcal I_{\mathcal R}$ is some suitable set whose elements uniquely identify the orbits of $\mathcal R$. The elements of $\mathcal I_{\mathcal E} $ and $\mathcal I_{\mathcal R} $ should be considered "orbit identifiers", as they are in a 1-1 correspondence with orbits in $\mathcal E$ and $\mathcal R$.
\subsubsection{Equivariance of the Information Loss Map and the Orbit Connector}
To make use of the developed group theoretic structures, the information loss map $\Omega$ must be compatible with the group actions.\\
The information loss map $\Omega$ is called \textbf{equivariant} if it satisfies
\begin{equation}
\Omega(T \star E)=T \star \Omega(E)
\end{equation}
meaning that a time change in an evolutionary process results in an equivalent time change of the associated reconstructed process (and vice versa).  \\
For the rest of this section I assume that $\Omega$ is equivariant. This is a natural property in probabilistic settings, where the time change is a reformulation of the pushforward measure or distribution and reflects the compatibility of the involved structures with preimages (see example).
\subsection{Hierarchies of Non-identifiability}
As a result of the equivariance, $\Omega$ maps orbits in $\mathcal E$ to orbits in $\mathcal R$. Thus ambiguity can occur at distinct scales:
\begin{itemize}
\item \textbf{Small scale identifiability/ambiguity}: Is there a 1-1 correspondence between evolutionary processes from an orbit $\mathcal O_E $ and the reconstructed process from its associated orbit $\Omega (\mathcal O_E):= \mathcal O_R $
\item \textbf{Intermediate scale identifiability/ambiguity}: Is there a 1-1 correspondence between orbits in $\mathcal E$ and orbits in $\mathcal R$?
\end{itemize}
Only if identifiability at both these scales holds, it can hold between all of $\mathcal E$ and $\mathcal R$.
\subsection{Identifiability and Ambiguity on the Small Scale}
Here, I provide necessary and sufficient conditions for small scale identifiability, meaning that there is a 1-1 correspondence between the evolutionary processes in an orbit $\mathcal O \subset \mathcal E$ and the reconstructed processes of its associated orbit $\Omega(\mathcal O) \in \mathcal R$.\\
The \textbf{isotropy group} of an evolutionary process $E$ under time change is given by
\begin{equation}
G_{E}:= \{ T \in \mathcal T \mid T \star E = E\}
\end{equation}
It contains all time scales that leave the process $E$ unchanged. Isotropy groups are subgroups of $\mathcal T$, and for all $E \in \mathcal O$, they come from the same conjugacy class called the \textbf{isotropy type} of $\mathcal O$ \cite{lee2010} . The isotropy group of a reconstructed process $R$ and the isotropy type of the orbit $\Omega( \mathcal O )$ are defined mutatis mutandis.\\
With these definitions, the central result on identifiability on the small scale is
\begin{itemize}
\item $\Omega$ is a 1-1 correspondence between $\mathcal O$ and $\Omega(\mathcal O)$ if and only if they have the same isotropy type \citep[p. 290]{lee2010}. 
\end{itemize}
As the isotropy groups of a free group action are trivial, a direct conclusion is that if $\mathcal T$ acts freely on both $\mathcal O$ and $\Omega(\mathcal O)$, then $\Omega$ provides an 1-1 correspondence between the two. Thus we get the following result:
\begin{itemize}
\item If $\mathcal T$ acts freely on both $\mathcal E$ and $\mathcal R$, then $\mathcal E$ is identifiable on the small scale.
\end{itemize}
\subsection{(Non)identifiability on the Intermediate Scale}
Here I provide a framework to examine identifiability and ambiguity on the intermediate scale.\\
As $\Omega$ maps orbits in $\mathcal E$ to orbits in $\mathcal R$, it gives rise to 
the \textbf{orbit connector} map
\begin{equation}
\omega \colon \mathcal I_{\mathcal E} \to  \mathcal I_{\mathcal R}
\end{equation}
that maps orbit identifiers of $\mathcal E$ to the identifiers of their associated orbits in $\mathcal R$. \ref{fig:invariants}. The orbit connector is a low resolution equivalent of $\Omega$ that is indifferent to changes in time scale and provides a complete description of the identifiability problem on the intermediate scale. By definition, the cardinality of the fiber $\omega^{-1}( \mathcal K_{\mathcal R}(\mathcal O_R))$ corresponds to the number of orbits in $\mathcal E$ that are mapped to the orbit $\mathcal O_R$.\\
This directly implies the following result:
\begin{itemize}
\item The identifiability problem on the intermediate scale is solvable if and only if $\omega$ is bijective.
\item The cardinality of the fibers of $\omega$ corresponds to the ambiguity on the intermediate scale.
\end{itemize}
To construct $\omega$ from the $\Omega$, it is helpful to introduce the \textbf{information loss invariant} (associated with $\mathcal K_{\mathcal R}$), a map
\begin{equation}
\mathcal K_{\Omega} \colon \mathcal E \to \mathcal I_{\mathcal R}
\end{equation}
defined as
\begin{equation}
\mathcal K_{\Omega}:= \mathcal K_{\mathcal R} \circ \Omega
\end{equation}
\begin{figure}
    \centering
\begin{tikzcd}
 \mathcal I_ \mathcal E \arrow[r , "\omega"] & \mathcal I _ \mathcal F \\
\mathcal E \arrow[ru, "K_\Omega"] \arrow[r, "\Omega"] \arrow[u, "\mathcal K_\mathcal E"] & \mathcal R \arrow[u, "\mathcal K _{\mathcal R}" ] 
\end{tikzcd}
\caption{Structure of the intermediate identifiability problem} \label{fig:invariants}
\end{figure}

It maps evolutionary processes to the (identifier of the) orbit in $\mathcal{R}$ that contains their associated reconstructed process. It is invariant because of the equivariance of $\Omega$ and the invariance of $\mathcal K_{\mathcal R}$, however it it in general not maximalinvariant.\\
If available, the information loss invariant can be factorized as
\begin{equation}
\mathcal K_{\Omega}=\omega \circ \mathcal K_{\mathcal E}
\end{equation}
from which information on $\omega$ can be recovered (see example).
\subsection{Sufficient Criteria for Distinguishability of Evolutionary Processes}
The information loss invariant provides a sufficient criterion for the distinguishability of evolutionary processes. This is based on the observation that evolutionary processes $E_1, E_2$ with $\mathcal K_\Omega(E_1) \neq \mathcal K_{\Omega}(E_2)$ are mapped to different orbits in $\mathcal R$, and their reconstructed processes are thus necessarily distinct:
\begin{itemize}
\item If evolutionary processes have different information loss invariants, then they come from different congruence classes and can thus be distinguished based on the reconstructed processes.
\end{itemize}
\subsection{Invariant and Equivariant Functions of Reconstructed Processes}
Here I examine the implications of the results on the identifiability of equivariant and invariant inferences from $\mathcal R$.
\subsubsection{Invariant Functions}
Let
\begin{equation}
 f^{\text{invar}}_{\mathcal R} \colon \mathcal R \to \mathcal F
\end{equation}
be an invariant function on $\mathcal R$, and
\begin{equation}
 f^{\text{invar}}_{\mathcal E} \colon \mathcal E \to \mathcal F
\end{equation}
defined as 
\begin{equation}
 f^{\text{invar}}_{\mathcal E} :=  f^{\text{invar}}_{\mathcal R} \circ \Omega 
\end{equation}
the corresponding (invariant) inference from the evolutionary processes.
As the reconstruction invariant $\mathcal K_{\mathcal R}$ is maximalinvariant, $f^{\text{invar}}_{\mathcal R}$ can be written as
\begin{equation}
f^{\text{invar}}_{\mathcal R} = G_{\mathcal R} \circ \mathcal K_{\mathcal R}
\end{equation}
for some function
\begin{equation}
G_{\mathcal R} \colon \mathcal I_{\mathcal R} \to \mathcal F
\end{equation}
Thus every invariant functions on $\mathcal R$ is a function of the reconstruction invariant. The inference from the evolutionary processes can be written as
\begin{equation}
 f^{\text{invar}}_{\mathcal E}= G_{\mathcal R} \circ \mathcal K_{\mathcal R} \circ \Omega =G_{\mathcal R} \circ \mathcal K_{\Omega} 
\end{equation}
showing that 
\begin{itemize}
\item Every invariant inference about evolutionary processes that is based on reconstructed processes is a function of the information loss invariant $\mathcal K_{\Omega}$. As such, it can only contain the same or less information than $\mathcal K_{\Omega}$.
\end{itemize}

\subsubsection{Equivariant Functions}
Let
\begin{equation}
 f^{\text{equi}}_{\mathcal R} \colon \mathcal R \to \mathcal F
\end{equation}
be an equivariant function on $\mathcal R$ and let
\begin{equation}
 f^{\text{equi}}_{\mathcal E} \colon \mathcal E \to \mathcal F
\end{equation}
defined as 
\begin{equation}
 f^{\text{equi}}_{\mathcal E} :=  f^{\text{equi}}_{\mathcal R} \circ \Omega 
\end{equation}
the corresponding inference from the evolutionary processes.\\
The identification of elements of $\mathcal R$ based on elements of $\mathcal F$ yields a secondary identification problem that is structurally identical to the primary one between $\mathcal R$ and $\mathcal E$ discussed above. Accordingly identifying the elements of $\mathcal E$ based on the elements of $\mathcal R$ stacks the difficulties of both identifiability problems:
\begin{itemize}
\item The small (resp. intermediate) scale identifiability problem between $\mathcal E$ and $\mathcal F$ is solvable if and only if both the small (resp. intermediate) scale identifiability problem between $\mathcal E$ and $\mathcal R$ and between $\mathcal R$ and $\mathcal F$ are solvable.
\end{itemize}
Thus
\begin{itemize}
\item Every equivariant inference about evolutionary processes that is based on reconstructed processes has the same or more severe identifiability issues than the original problem of identifying evolutionary processes based on reconstructed processes. This holds both on the small and the intermediate scale.
\end{itemize}
\subsection{Summary of the Group Theoretic Approach}
\begin{itemize}
\item \textbf{Structural Origins}:The identifiability problem can be separated in two scales:
\begin{enumerate}
\item \textbf{Small scale}: Changes in time scale determine identifiability
\item \textbf{Intermediate scale}: The Invariants under changes in time scale determine identifiability.
\end{enumerate} The identifiability problems on these scales determine the properties of the identifiability problem in the large scale.
\item \textbf{Distinguishability}: The information loss invariant provides a sufficient criterion for distinguishability.
\item \textbf{Propagation}: When inferring about evolutionary processes based on reconstructed processes:
\begin{enumerate}
\item Any inference that is not affected by changes in time scale is a function of the information loss invariant.
\item Any inference about temporal processes that is compatible with changes in time scale has the same or more severe identifiability issues as the original problem.
\end{enumerate}
\item \textbf{Dissimilarity}: If different orbits contain evolutionary processes with distinct characteristics, then non-identifiability on the intermediate scale leads to the loss of these distinct characteristics
\end{itemize}
\section{Example: Identifiability of Birth-Death Processes} \label{birthdeath}
In this example, I demonstrate the tools developed above by applying them to the birth-death processes (evolutionary processes) and the reconstructed processes (sensu \citep{nee1994b}).\\ 
Let $\mathcal Q(a,b)$ be the set of all strictly increasing continuously differentiable functions that are bijective from $[0,a]$ to $[0,b]$ and have $F(0)=0$ and $F(a)=b$. I use the notation $W|_{[a,b]}$ to indicate that the function $W$ has domain of definition $[a,b]$\\
I assume the processes start at time $t=0$ and are reconstructed at time $t^*>0$, so the time interval of interest is $I=[0,t^*]$. A birth-death process is uniquely determined by its birth rate $\mu(t)$ and death rate $\mu(t)$, which I assume to be continuous and strictly positive. To obtain a simpler representation of time change, I consider the integrated birth rate
\begin{equation}
\Lambda(t):= \int_0^{t} \lambda(x) \mathrm dx \in \mathcal Q(t^*,\Lambda(t^*))
\end{equation}
and the integrated death rate
\begin{equation}
M(t):= \int_0^{t} \mu(x) \mathrm  dx \in \mathcal Q(t^*,M(t^*))
\end{equation}
instead on $\mu(t)$ and $\lambda(t)$. With this, I will identify any evolutionary process by its associated pair $(M, \Lambda)$.\\
A reconstructed process in uniquely determined by its diversification rate $\delta(t)$, which I assume to be strictly positive and continuous. As above, I use
\begin{equation}
\Delta(t):= \int_0^t \delta(x) \mathrm dx \in \mathcal Q(t^*,\Delta(t^*))
\end{equation}
instead, and any reconstructed process is uniquely identified by its $\Delta$.
\subsubsection{Group structure}
The group of time changes is given by $\mathcal T = \mathcal Q(t^*, t^*)$. The group operation is the composition of functions, the inverse of $T$ is given by its inverse function $T^{-1}$, thus the identity is the function $T=\operatorname{Id}|_{[0,t^*]}$.\\
The group $\mathcal T$ acts on $\mathcal R$ via
\begin{equation}
T \star \Delta := \Delta \circ T^{-1}
\end{equation}
and on $\mathcal E$ via
\begin{equation}
T \star (M, \Lambda) := (M \circ T^{-1}, \Lambda \circ T^{-1})
\end{equation}
The orbits in $\mathcal E$ are given by
\begin{equation}
\mathcal O_{(M,\Lambda)}= \{ (M', \Lambda') \mid (M', \Lambda') = (M \circ T^{-1}, \Lambda \circ T^{-1}) \text{ for some } T \in \mathcal T \}
\end{equation}
and the orbits in $\mathcal R$ by
\begin{equation}
\mathcal O_{\Delta}= \{ \Delta' \mid \Delta' = \Delta \circ T^{-1} \text{ for some } T \in \mathcal T \}
\end{equation}

\subsection{Equivariance of $\Omega$}
The equivariance of $\Omega$ is a direct result of the fact that the involved processes can  be taken as (random or deterministic) measures, and the equivariance is equivalent to their compatibility with the image measure. More specifically, $\Omega$ by definition satisfies
\begin{equation}
    \Omega((M(t), \Lambda(t))) = \Delta(t)
\end{equation}
for all $t$. But then it also satisfies
\begin{equation}
    \Omega((M(T^{-1}(t')), \Lambda(T^{-1}(t')))) = \Delta(T^{-1}(t'))
\end{equation}
for all $t'$, which shows equivariance of $\Omega$. Note that this argument does not require any specific definition of $\Omega$, it simply uses the fact that connects two time-integrated quantities (measures on the time axis) with each other.\\

\subsection{Identifiability on the Small Scale}
With the equivariance of $\Omega$ in place, we can already address identifiability on the small scale. The group action $\star$ is free on both the evolutionary and the reconstructed processes. As a result, $\mathcal E$ is identifiable on the small scale: $\Omega$ is a 1-1 correspondence between birth-death processes in an orbit $\mathcal O_E$ and the reconstructed processes in the associated orbit $\mathcal O_R = \Omega(\mathcal{O}_E)$.\\
The fact that the group action is free is rooted in the bijectivity of the accumulated rates that characterize the involved processes. It is notable that changing the model assumptions to include rates that can be zero over short intervals would result in a group action that is not free, which will introduce identifiability issues on the small scale.

\subsection{Identifiability on the Intermediate Scale}
To examine the identifiability on the intermediate scale, I first define the reconstruction and evolutionary invariants.

\subsubsection{Reconstruction Invariant}
As reconstruction invariant on $\mathcal R$, I use
\begin{equation}
\mathcal K_{\mathcal R} \colon \mathcal R \to (0,\infty)
\end{equation}
defined by
\begin{equation}
\mathcal K_{\mathcal R}(\Delta)= \Delta(t^*)
\end{equation}
It assigns each reconstructed process the accumulated diversification rate it has amassed up to the time of observation $t^*$. The unique orbit $\mathcal O_r$ associated with $r \in (0,\infty)$ is
\begin{equation}
\mathcal O_r = \mathcal K_{\mathcal R}^{-1}(r)=\mathcal Q(t^*, r)
\end{equation}
\subsubsection{Evolutionary Invariant}

Define
\begin{equation}
\mathcal I_{\mathcal E} := \{ F \mid F \in \mathcal Q(a,b) \text{ for some } a,b > 0 \}
\end{equation}
As evolutionary invariant
\begin{equation}
\mathcal K_{\mathcal E} \colon \mathcal E \to \mathcal I_{\mathcal E}
\end{equation}
I use
\begin{equation}
\mathcal K_{\mathcal E}((M, \Lambda)):=\Lambda \circ M^{-1} \quad ( \in \mathcal Q(M(t^*), \Lambda(t^*))
\end{equation}
It assigns values of the accumulated death rate to the values of the corresponding accumulated birth rate. Thus, it determines how much birth rate a process has "experienced" when its accumulated death rate has reached a certain value. This is an evolutionary measure that connects the exposure to extinction risk the process has experienced with the amount of speciation it has displayed until this exposure.\\
The orbit $\mathcal O_W$ associated with a function $W \in \mathcal I_{\mathcal E}$ defined on $[0,a]$ is given by
\begin{equation}
\mathcal O_W = \mathcal K_{\mathcal E}^{-1}(W|_{[0,a]})= \{ (M,W \circ M) \mid M \in \mathcal Q(t^*,a)  \}
\end{equation}
Both invariants together already hint at an underlying identifiability issue. Recall that identifiability on the intermediate scale holds if there is a bijection $\omega$ between the evolutionary and reconstruction invariants. In the examined case, the reconstruction invariants are simple (an interval of real numbers), while the evolutionary invariants are much more complex and appear to be of a larger cardinality as they are a subset of a function space. While it is hard to show that there is no bijection between them, this discrepancy in size and complexity points to an underlying tension.

\subsection{Identifiability on the Intermediate Scale}
\subsubsection{Information Loss Invariant}
The information loss invariant associated with $\mathcal K_{\mathcal R}$ is
\begin{equation}
\mathcal K_{\Omega} \colon \mathcal E \to \mathcal I_{\mathcal R}
\end{equation} 
defined as
\begin{equation}
\mathcal K_\Omega((\Lambda, M)):= \ln \left( 1+ \int_0^{t^*}   M'(t) \exp(M(t)-\Lambda(t)) \mathrm dt \right)
\end{equation}
It assigns each birth-death process the diversity of its associated reconstructed process at the time of reconstruction $t^*$.\\
For intermediate identifiability, I characterize all evolutionary invariants that are mapped on the same reconstruction invariant $r$.
\subsubsection{Orbit Connector Map}
Integration by substitution with $M^{-1}$ in $\mathcal K_\Omega$ shows that the orbit connector
\begin{equation}
\omega \colon \mathcal I_{\mathcal E} \to \mathcal I_{\mathcal R}
\end{equation}
is given by
\begin{equation}
\omega(W) = \ln \left( 1+ \int_0^a \exp(x-W(x)) \mathrm dx \right)
\end{equation}
for any $W=W|_{[0,a]} \in \mathcal I_{\mathcal E}$.\\
The fibers of $\omega$ determine the degree of ambiguity of the reconstruction. To characterize them, define 
\begin{equation}
\Xi_r := \{f \mid f \in \mathcal Q(a,\exp(r)-1) \text{ for some } a \text{ and } \tfrac{f^{''}}{f'} < 1\}
\end{equation}
Examples for functions in $\Xi_r$ are $\exp(cx)-1$ for $c<1$ and any $r>0$. The fibers of $\omega$ are uniquely characterized by
\begin{align}
G \in \omega^{-1}(r) \iff
 G = \operatorname{Id} - \ln \circ f'|_{[0,f^{-1}(\exp(r)-1)]} \text{ for some } f \in \Xi_r
\end{align}
\subsection{Invariant and Equivariant Inferences}
Based on the characterization of invariant inferences in the section above, every invariant inference is a function of the reconstruction invariant and thus of the integrated diversification rate at the time of reconstruction $t^*$\\
Based on the identifiability on the small scale and results for the ambiguity on the intermediate scale, the statement for equivariant inferences can made more precise:
\begin{itemize}
\item Given we want to infer about a temporal process in evolutionary processes based on information about reconstructed processes, and that temporal process is compatible with time change. Then there are uncountably many evolutionary processes that generate the same temporal process.
\end{itemize}
\subsection{Summary Identifiability of Birth-Death-processes}
\begin{itemize}
\item \textbf{Ambiguity}: Uncountably many birth-death processes are associated with the same reconstructed process.
\item \textbf{Propagation:} When inferring about evolutionary processes based on reconstructed processes.
\begin{enumerate}
\item Any inference that is not affected by changes in time scale is a function of integrated diversification rate at the time of observation.
\item Any inference about processes through time that is compatible with time change is ambiguous, and uncountably many evolutionary processes are mapped to the same process through time.
\end{enumerate} 
\item \textbf{Distinguishability:} The information loss invariant provides a sufficient criterion for distinguishability of evolutionary processes.
\item \textbf{Dissimilarity:} Evolutionary processes with distinct evolutionary dynamics are mapped to the same reconstructed process, leading to a loss of this information.
\item \textbf{Structural Origins}: The model assumption that rates are nonzero ensures that the small scale identifiability problem is uniquely solvable as the group action is free. Weakening this assumption will introduce additional identifiability issues.
\end{itemize}
\section{Topological Invariants and (Non)Identifiabiltiy} \label{topology}
In this section, I examine how topological invariants of $\mathcal E$ and $\mathcal R$ control the degree of non-identifiability.
\subsection{Preliminaries and Assumptions}
To have a minimal topological framework, let $\mathcal E$ and $\mathcal R$ be equipped with some topologies, and let $\Omega$ be continuous with respect to these topologies.\\
I assume that
\begin{itemize}
\item $\mathcal E$ is connected, meaning that the evolutionary processes form a topological unit that can not be separated into two or more disjoint open sets.
\item $\mathcal E$ is locally path connected, meaning that any two evolutionary processes $E_1, E_2$ close enough to a evolutionary process $E$ can be connected by a continuous path.
\end{itemize}
If $\mathcal E$ fails to be connected, the examination of identifiability issues can be continued on its components that form independent subdivisions of $\mathcal E$. Path connectivity can be thought of as a type of local continuity requirement, expressed as the ability to continuously morph evolutionary processes into each other given they are sufficiently close to each other.\\
For $\Omega$, I assume
\begin{itemize}
\item \textbf{Local identifiability}: Every evolutionary process $E$ has a connected open neighborhood $U_E$ such that $\Omega$ is a local homeomorphism from $U_E$ to $U_R:=\Omega(U_E)$. This means that at least locally, $\Omega$ provides a continuous 1-1 connection between evolutionary processes and reconstructed processes.
\item \textbf{Separability} in congruence classes: For all evolutionary processe $E$ from the same congruence class $\mathbf C_R$, the neighborhoods $U_E$ can be chosen to be disjoint.
\end{itemize}
Naturally, local identifiability is a necessary condition for the global identifiability that is of interest here. The separability of congruence classes ensures its individual elements can be distinguished properly.
\subsection{Information Loss as Covering Map}
When the assumptions given above hold, $\Omega$ becomes a covering map from $\mathcal E$ to $\mathcal R$. Here, each of the neighborhoods $U_E$ of the congruence classes "covers" the neighborhood $U_R$ of the associated reconstructed process.\\
A central result for covering maps is that
\begin{itemize}
\item All congruence classes have the same cardinality (\citep{lee2010} p. 279)
\end{itemize}
Thus, the difficulty of the identifiability problem is not dependent on the choice of reconstructed processes.
\subsection{Information Loss and the Fundamental Group}
Homotopy theory provides a tight connection between the properties of covering maps and the fundamental groups of $\mathcal E$ and $\mathcal R$ that  theory provides further insights into the size of congruence classes.\\
Recall that a topological space is called simply connected if every loop can be continuously shrunk to a point. It can be shown that (See \citep[p. 292]{lee2010} )
\begin{itemize}
\item If $\mathcal E$ is simply connected, then the congruence classes have the same number of elements as the fundamental group of $\mathcal R$.
\item If $\mathcal R$ is simply connected, all evolutionary processes can be uniquely identified
\end{itemize}
\subsection{Complexity of Models Relative to Each Other}
Here I provide a general framework in which the above results fit and that shows that the ambiguity is determined by the complexity of the models relative to each other. For this section, let $\pi_1(\mathcal R,R_0)$ be the fundamental group of $\mathcal R$ \\
An automorphism on $\mathcal E$ is a bijective continuous map with continuous inverse from $\mathcal E$ to itself. With the  composition of functions and the inverse functions, automorphisms form a group $\operatorname{Aut}(\mathcal E)$.
This group is tightly connected with the identifiability, as automorphisms permute the elements of congruence classes, which can be used to infer about their properties and size. The main result is (cf. \citep{lee2010}, p. 310)
\begin{itemize}
\item If $\Omega$ is a normal covering, then $\operatorname{Aut}_\Omega(\mathcal E)$ is isomorphic to the quotient of $\pi_1(\mathcal R, R_0)$ and the subgroup $\Omega_*\pi_1(\mathcal E, E_0)$ induced by $\Omega$ 
\end{itemize}
Note that for normal coverings, the automorphism group acts freely and transitively on congruence classes (\citep{lee2010} p. 210), thus they have the same cardinality according to the orbit-stabilizer theorem.\\
The quotient group in the above theorem provides a measure for the complexity of $\mathcal E$ relative to $\mathcal R$, with the limiting cases (compare results above)
\begin{itemize}
\item Maximum relative complexity: If  $\mathcal E$ is simply connected, then its induced subgroup is the identity, and the quotient is $\pi_1(\mathcal R, R_0)$
\item Minimum relative complexity: If  $\mathcal R$ is simply connected, then $\Omega$ is a homeomorphism and provides a continuous and invertible 1-1 correspondence between the reconstructed and the evolutionary processes.
\end{itemize}
\subsection{Adjusting Models to Data Availability}
I examine the mitigation strategy of reducing the identifiability issues by changing the evolutionary processes whilst data availability remains the same. For this, I take the reconstructed processes as models of available data.\\
Within the framework of covering maps, this mitigation strategy is equivalent of adjusting $\mathcal E$ and $\Omega$ such that $\Omega$ remains a covering map, i.e. still allows local identifiability.\\
For this question, I assume $\mathcal R$ has a universal covering space, which is the case if it is connected and locally simply connected (\citep{lee2010} theorem 11.43). The the following statement holds:
\begin{itemize}
\item There is a 1-1 correspondence between (1) conjugacy classes of subgroups of $\pi_1(\mathcal R,R_0)$ and (2) isomorphism classes of coverings of $\mathcal R$ (\citep{lee2010} p. 315)
\end{itemize}
It associates a covering of $\mathcal R$ with the conjugacy class of its induced subgroup. Thus, the subgroups of the fundamental group of the reconstructed processes determine what coverings can exist. This shows that size of congruence classes (and thus the degree of ambiguity) can not be improved on arbitrarily whilst preserving local identifiability.
\begin{itemize}
\item The subgroups of the fundamental group of the reconstructed processes $\mathcal R$ determine what degrees of ambiguity can occur when changing the evolutionary processes $\mathcal E$.
\end{itemize}
\subsection{Summary of the Topological Approach}
\begin{itemize}
\item \textbf{Ambiguity:} The identifiability problem is equally difficult everywhere, meaning the ambiguity is independent of the choice of reconstructed process.
\item \textbf{Structural Origins:} The difficulty of the identifiability problem is determined by the topological complexity of the reconstructed processes relative to the complexity of the evolutionary processes. This can be made explicit using their fundamental groups and their induced subgroups
\item \textbf{Ambiguity and Identifiability:} End-members of the above statement are:
\begin{enumerate}
\item If the reconstructed processes are topologically simple (i.e. simply connected), all evolutionary processes can be identified based on reconstructed processes
\item If the evolutionary processes are topologically simple, then the degree of ambiguity is determined by the topological complexity of the reconstructed processes.
\end{enumerate}  
\item \textbf{Mitigation:} For a fixed amount of data, the ambiguity regarding evolutionary processes can not change arbitrarily, but is constrained  by the topological structure of the reconstructed processes. This limits the models of evolutionary processes that can be used to match available data.
\end{itemize}

\bibliography{lib.bib} 
\bibliographystyle{abbrvnat}
\end{document}